\documentclass[a4paper,11pt]{article}

\usepackage{amsmath,amssymb,color,graphics,epsfig,cite}
\usepackage{titlesec}
\usepackage[titletoc,title]{appendix}

\usepackage{amsfonts}

\newcommand{\hoch}[1]{$\, ^{#1}$}

\newcommand{\be}{\begin{equation}}
	\newcommand{\ee}{\end{equation}}
\newcommand{\bea}{\setlength\arraycolsep{2pt} \begin{eqnarray}}
	\newcommand{\eea}{\end{eqnarray}}
\newcommand{\nn}{\nonumber}

\def\ft#1#2{{\textstyle{\frac{\scriptstyle #1}{\scriptstyle #2} } }}
\def\fft#1#2{{\frac{#1}{#2}}}

\def\0{{\sst{(0)}}}
\def\1{{\sst{(1)}}}
\def\2{{\sst{(2)}}}
\def\3{{\sst{(3)}}}
\def\4{{\sst{(4)}}}
\def\5{{\sst{(5)}}}
\def\6{{\sst{(6)}}}
\def\7{{\sst{(7)}}}
\def\8{{\sst{(8)}}}
\def\sst#1{{\scriptscriptstyle #1}}

\begin{document}
	
	\begin{center}
		{\Large {\bf Weak Cosmic Censorship Conjecture Cannot be Violated\\ in Gedanken Experiments}}
		
		\vspace{20pt}
		
		Peng-Yu Wu\hoch{1}, H. Khodabakhshi\hoch{1*} and H. L\"{u}\hoch{1,2}
		
		\vspace{10pt}
		
		{\it \textsuperscript{1}Center for Joint Quantum Studies and Department of Physics,\\
			School of Science, Tianjin University, Tianjin 300350, China }
		
		\bigskip
		
		{\it \textsuperscript{2}Joint School of National University of Singapore and Tianjin University,\\
			International Campus of Tianjin University, Binhai New City, Fuzhou 350207, China}
		
		\vspace{40pt}
	\end{center}
	
	\begin{abstract}
		We innovate a systematic investigation of the Weak Cosmic Censorship Conjecture (WCCC) using gedanken experiments involving black hole perturbations by test particles. We classify various WCCC violation scenarios proposed in recent decades, including Hubeny, mixed, and the latest Sorce-Wald (SW). We provide general formulae in each case, and resolve contradictions in numerous studies. Following SW type, our analysis reveals that WCCC depends on the sign choice of the parameter $W\equiv \big(\frac{\partial S}{\partial T}\big)_{Q_i; T=0}$ of the extremal black holes. $W > 0$ preserves WCCC and $W < 0$ would indicate potential violation. We show explicitly that $W>0$ for spherically-symmetric and static black holes, and for general case, we argue that it is protected by the black hole no-hair theorem. We also consider asymptotically-(A)dS black holes and argue that there can be no violation either.
	\end{abstract}
\vfill

{\footnotesize \hoch{*}Corresponding author (h\_khodabakhshi@tju.edu.cn)}

\thispagestyle{empty}
\pagebreak

\tableofcontents
\addtocontents{toc}{\protect\setcounter{tocdepth}{2}}

	\section{Introduction}
	
	The Weak Cosmic Censorship Conjecture (WCCC), proposed by Penrose \cite{Penrose}, states that for a geodesically complete and asymptotically flat spacetime, the evolution of matter fields satisfying the null energy condition (NEC) cannot lead to a naked singularity. The importance of this conjecture cannot be overestimated since it preserves the predictive power of physical laws by preventing a visible singularity from being observed at infinity through physical processes \cite{Tsingh}. The general proof or demonstration of WCCC is challenging, one approach involves perturbing a black hole to determine whether there exists a critical situation that can lead to a visible singularity. According to the gedanken experiments proposed by Wald \cite{Wald}, by throwing or dropping matter, carrying energy $E > \delta M$ and charges $q_i = \delta Q_i$, into a black hole of $(M, Q_i)$, the final state will have mass $M + \delta M$ and charges $Q_i + \delta Q_i$, where the matter obeys the null energy condition (NEC). Note that $Q_i$ denote conserved quantities such as electric charges and/or angular momentum, {\it etc.} Applying the thermodynamic second law $\delta S \ge 0$, the linear variation of black hole mechanics leads to
	\begin{equation}\label{NEC}
		\delta M \ge \phi_i \delta Q_i\,.
	\end{equation}
Note that the extremal case $(T=0)$ always saturates the inequality. For asymptotically Minkowski spacetimes, the null energy condition (NEC) implies that the flux of energy and charges must be positive, leading to inequality \eqref{NEC} \cite{Poisson,Ht1}. This inequality can also be derived for an infalling particle in geodesic calculations, e.g., in the case of a Kerr-Newman (KN) black hole \cite{Wald,Zhang}. Focusing on the KN solution, Wald proposed that if the horizon condition, characterized by
	$
	M^2 \ge Q^2 + \left(J/M\right)^2,
	$
	were not satisfied after perturbation, the spacetime would contain a naked singularity \cite{Wald}. He showed that no violation of WCCC could occurred by throwing particle matter into an extremal KN black hole. However, Hubeny and Jacobson \cite{Hubney,Jacobson} suggested that the WCCC could be violated for a slightly non-extremal black hole. (We call it as Hubney's type.) Following these works, it has been shown \cite{Zhang} that WCCC can be violated for an extremal KN black hole by expanding the horizon condition up to the second order in $\delta M$ and $\delta Q_i$. (We call it as mixed type.) See \cite{Ht1, Ht2, Ht3,Yang1,Yang2} for some followup works on the Hubeny's type and \cite{Mt1, Mt2, Mt3, Mt4, Mt5,Sanjar0,Yang1,Yang2} on the mixed type.
	
	To resolve these contradictions with Hubeny or others, Sorce and Wald (SW) \cite{Wald1,Wald2} realized that the analysis in Hubeny's type of gedanken experiments \cite{Hubney} is insufficient at the linear order of $\delta Q_i$ and $\delta M$, while the near-extremal expansion is second order \cite{Sanjar1,Sanjar2}. We also find that there is confusion in mixing different orders in the mixed type analysis \cite{Zhang}. We shall clarify these points in this paper by investigating the problems and providing general formulae.
	
	SW proposed a method to examine the WCCC of the near extremal KN black hole for up to the second-order variation \cite{Wald1}. (We refer to this as the SW type.) At the second order, they suggested the following inequality
	\begin{equation}\label{w1}
		\delta^2 M - \Omega_H \delta^2 J - \phi_H \delta^2 Q \geq -\frac{k}{8\pi} \delta^2 A^{KN},
	\end{equation}
	which is the second-order generalization of black hole dynamics, (where the first-order variation is given by
	$
	\delta M \ge \phi_i \delta Q_i,
	$)
	and $\delta^2 A^{KN}$, as expressed in Eq.~(113) in \cite{Wald1}, is in fact the Hessian metric of $A^{KN}$. For the KN black hole, the existence horizon condition is \( f(\lambda) \ge 0 \), where
	\begin{equation}\label{w8}
		f(\lambda) = M(\lambda)^2 - Q(\lambda)^2 - \frac{J(\lambda)^2}{M(\lambda)^2}\,.
	\end{equation}
	Using the inequality (\ref{w1}) and \(\delta M = \phi_i \delta Q_i\), expanding the above equation up to the second order in \(\lambda\) and \(\epsilon\) leads to
\bea
f(\lambda) \geq \left(-\frac{(M^4-J^2) Q \delta Q + 2 J M^2 \delta J}{M(M^4 + J^2)} \lambda + M \epsilon\right)^2
+ O(\lambda^3, \lambda^2 \epsilon, \ldots)\,,\label{wald1}
\eea
	in which \(\epsilon = \frac{r_+}{M} - 1\ge0\) is a dimensionless blackening parameter. Note that at the above order, the mass $M$ should be taken to be the extremal value $M_{\rm ext}=\sqrt{\fft12(Q^2 + \sqrt{Q^4 + 4J^2})}$.
	
	In the gedanken experiment, we need to deal with two independent inequalities: one is the {\it energy condition} (it is equivalent to Eqs.~(\ref{NEC}) and (\ref{w1}), up to the first and second orders, respectively), and the other is the existence of horizon condition, or {\it horizon condition} for short. In other words, $M > M_{\text{ext}}$ must be satisfied before and after perturbation to protect WCCC. In the perturbation process, we therefore need to be careful about how to expand the two inequality equations in terms of small quantities up to the appropriate orders, and we need to avoid mixing different orders in the near-extremal limit. We shall clarify these points with great care.
	
	We first need to carefully define the meaning of \(\delta\) and \(\delta^2\) terms in the SW method  \cite{Wald1,Wald2}. Considering a function \(F[Q(\lambda)]\), the Taylor expansion up to the second order at \(\lambda = 0\) is
	$
	F[Q(\lambda)] = F[Q(0)] + F'[Q(0)] Q'(0) \lambda + \frac{1}{2} \left(F''[Q(0)] Q'(0)^2 + F'[Q(0)] Q''(0)\right) \lambda^2 + \cdots.
	$
	Alternatively, expanding
	$
	F[Q + \lambda \delta Q + \frac{1}{2} \lambda^2 \delta^2 Q],
	$
	up to the second order in \(\lambda\) gives the same result, where \(Q=Q(0)\), \(\delta Q = Q'(0)\) and \(\delta^2 Q = Q''(0)\). It should be noted that expanding \(F[Q + \lambda \delta Q]\) is different from $	F[Q(\lambda)]$ in the second order and does not include the \(Q''[0] = \delta^2 Q\) term. Thus, SW's second-order perturbation is based on this: if a particle with energy \(E > \Delta M\) and charge \(q_i = \Delta Q_i\) falls into a black hole, then the black hole's parameters will change to
	\begin{eqnarray}
		M &\to& M + \Delta M = M + \lambda \delta M + \frac{1}{2} \lambda^2 \delta^2 M + \cdots, \\
		Q_i &\to& Q_i + \Delta Q_i = Q_i + \lambda \delta Q_i + \frac{1}{2} \lambda^2 \delta^2 Q_i + \cdots.
	\end{eqnarray}
	
The main focus of this paper is to generalize the \textit{horizon condition} after an infalling test particle perturbs a black hole. To achieve this, for the near-extremal case \(M(T_\epsilon, Q_i)\), we introduce the quantity \(X_{\epsilon}\), defined as
	\begin{equation}\label{M1}
			X_{\epsilon} \equiv M(T_\epsilon, Q_i) + \Delta M - M_{\text{ext}}(Q_i + \Delta Q_i) = \Big(M(T_\epsilon, Q_i) - M_{\text{ext}}(Q_i)\Big) + \Delta M - \Delta M_{\text{ext}}\,.
	\end{equation}
The sign of this quantity indicates whether the black hole's mass after perturbation, denoted as \(M_{\text{new}} = M(T_\epsilon, Q_i) + \Delta M\), remains larger than the extremal mass \(M_{\text{ext}}(Q_i + \Delta Q_i)\), where \(\Delta M\) is restricted by the \textit{energy condition}. In our analysis, \(X_{\epsilon}\) is directly related to the stability of the black hole's horizon. If the mass of the black hole is less than the extremal mass after perturbation (\(X_{\epsilon} < 0\)), it means the black hole is destroyed, or the WCCC can be violated. In the extreme case, we start from an extremal black hole where \(M_{\text{ext}}(Q_i) = M(T=0, Q_i)\), so that one can write
	\begin{equation} \label{M2}
		X_{\text{ext}} \equiv \Delta M - \Delta M_{\text{ext}}\,.
	\end{equation}
	Our conclusion is that WCCC cannot be violated in this gedanken experiment at the second order, owing to the fact that extremal black holes satisfy $W\ge 0$, with $W$ defined by
	\be
	W = \lim_{T\rightarrow 0} \Big(\fft{\partial S}{\partial T}\Big)_{Q_i} \equiv \Big(\fft{\partial S}{\partial T}\Big)_{Q_i;T=0}\,.\label{wdef}
	\ee
	In other words, for fixed charges, the extremal black hole has the minimum entropy.
	
	
	\section{Categorizing different types of WCCC}
	
	In this section, we use the proposed formulae in Eqs.~(\ref{M1}) and (\ref{M2}) to calculate the general formulae, critiquing the different well-known types of WCCC, including Hubeny's type, mixed type, and SW's type. In the last case, we first prove the Wald inequality (\ref{w1}) by using a new approach and considering $\delta S=0$ and $\delta^2 S>0$ \cite{Ningbo} in the second order of expansion. Then we show that for examining WCCC, we only need to look at the sign of the quantity $W$ for both the extreme and near-extreme cases.
	
	\subsection{Hubeny's type}
	
	The Hubeny's test of WCCC is effectively replacing $\Delta$ in \eqref{M1} by $\delta$, namely
	\begin{align}\label{H1}
		X_{\rm H}= \Big(M(T_\epsilon, Q_i) - M_{\text{ext}}(Q_i)\Big) + \delta M - \delta M_{\text{ext}}\,.
	\end{align}
	We perform the expansion near the extremal case up to the first order of \(\lambda\), where \(\delta^2 M = \delta^2 Q_i = 0\). Making use of the fact that
	$\delta M = T_{\epsilon} \delta S + \phi_{\epsilon,i} \delta Q_i$ and
	$\delta M_{\rm ext} = \phi_{\text{ext},i} \delta Q_i$, we find that near the extreme limit \(T = T_{\epsilon}\), we have
	\begin{eqnarray}\label{hmain}
	X_{\text{H}} = T_{\epsilon} \delta S + \ft{1}{2} W\, T_{\epsilon}^2 +  \lambda (\phi_{\epsilon,i} - \phi_{\text{ext},i}) \delta Q_i  \ge \ft12 W\, T_{\epsilon}^2 - \lambda\fft{\partial S_{\rm ext}}{\partial Q_i} \delta Q_i\, T_{\epsilon}\,,
\end{eqnarray}
	where the inequality arises from \(\delta S \ge 0\) and $W$ is define by \eqref{wdef}. The last term above can also be simply expressed as $T_{\epsilon} \delta S_{\rm ext}$. In the above derivation, we used the cyclic differential identity \be
	\Big(\fft{\partial x}{\partial y}\Big)_z\, \Big(\fft{\partial y}{\partial z}\Big)_x\, \Big(\fft{\partial z}{\partial x}\Big)_y=-1\,.\label{cyclicid}
	\ee
	The bracket term in \eqref{H1} occurs frequently in this paper and it is given by
\begin{eqnarray}
	M(T_\epsilon, Q_i) - M_{\text{ext}}(Q_i) = M(T_\epsilon, Q_i) - M(0,Q_i)=
	\ft{1}{2} W\, T_{\epsilon}^2 + {\cal O}(T_\epsilon^3)\,.\label{e1}
\end{eqnarray}
	Note that in this paper, $\epsilon$ denotes some non-extremal (dimensionless) parameter with $T_\epsilon \propto \epsilon + {\cal O}(\epsilon^2)$ as $\epsilon\rightarrow 0$. In the extremal limit, $T_{\epsilon}=0$, we simply have $X_{\rm H}=0$. In fact, our general formula (\ref{hmain}) summarizes all results of previous papers that adopted Hubeny's method, as in \cite{Ht1,Ht2,Ht3,Yang1,Yang2}. For example, for the RN black hole we have $T_{\epsilon} = \frac{\epsilon}{2 \pi M}$ and $S_{\rm ext}= \pi Q^2$, then one can obtain \(X_{\rm H} \ge \frac{M \epsilon^2}{2} - \lambda\epsilon \delta Q\) \cite{Hubney}. (For the more complicated KN black hole \cite{Ht3} see Appendix A.1.)  As is obvious, the first term has the \(\epsilon^2\) order, and the second term has \(\lambda \epsilon\) order (since \(\delta Q\) carries \(\lambda\) order). As we have discussed, the appropriate orders should have also include the \(\lambda^2\) term; therefore, Hubeny's method is incomplete.
	
	\subsection{Mixed type}
	
	 Another type of investigating WCCC is related to the mixing of orders in the extremal case, where \(\epsilon = 0\) and we have Eq.~(\ref{M2}). In \cite{Zhang,Mt1,Mt2,Mt3,Mt4,Mt5,Sanjar0,Yang1,Yang2}, the expansions up to the second order of \(\lambda\) have been performed for various explicit examples, but with \(\delta^2 M = \delta^2 Q = 0\). Their results can be summarized in the following simple universal formula
	\begin{equation}\label{main0}
		X_{\text{ext,mix}} = -\frac{1}{2} \lambda^2 \frac{\partial^2 M_{\text{ext}}}{\partial Q_i \partial Q_j} \delta Q_i \delta Q_j + \cdots\,,
	\end{equation}
	where we used \(\delta M = \phi_{\text{ext},i} \delta Q_i\) and \(T = 0\) in the extremal case, expanding \(\delta M_{\text{ext}}\) up to the second order of \(\lambda\). The result is simply the Hessian metric of extremal mass in terms of charges. Our simple equation (\ref{main0}) summarizes the results of several papers. For example, see Appendix A.2 for the KN black hole in \cite{Zhang}. This illustrates the power and efficiency of our method to clarify the problem by providing a general formula instead of going through the analysis for each specific black hole. However, as is obvious, in this mixed type, we performed the expansion for \(\delta M\) and \(\delta M_{\text{ext}}\) up to different orders; therefore, the sign of Eq.~(\ref{main0}) is insufficient to determine whether the WCCC can be violated.
	
	\subsection{SW type}
	
	 Before we derive the general formula for \(X_{\epsilon}\) in the appropriate order that includes all the \(\epsilon^2\), \(\epsilon \lambda\) and \(\lambda^2\) terms, proposed by SW for the KN black hole example, we would first like to understand the meaning of the inequality (\ref{w1}). It is instructive to calculate first
	\begin{align}\label{w4}
		S[\lambda] - S[M(\lambda), Q_i(\lambda)] = 0\,,
		M[\lambda] - M[S(\lambda), Q_i(\lambda)] = 0\,,
	\end{align}
	up to the second order in \(\lambda\), where the zeroth order is \(S[0] = S[M(0), Q(0)]\) or \(M[0] = M[S(0), Q(0)]\), corresponding to a thermodynamic system whose potential is $S$ or $M$ respectively. The first order expressions,
	\begin{align}
		\delta S = -\frac{\phi_i}{T} \delta Q_i + \frac{1}{T} \delta M\,,\qquad \delta M = T \delta S + \phi_i \delta Q_i\,,\label{w5}
	\end{align}
	are equivalent to the first law, and at the extreme case (\(T = 0\)), we have \(\delta M = \phi_{\text{ext},i} \delta Q_i\). The second order is
	\begin{align}\label{w6}
		&\delta^2 S = \frac{1}{T} \delta^2 M - \frac{\phi_i}{T} \delta^2 Q_i + DS,\nn \\
		&\delta^2 M = T \delta^2 S + \phi_i \delta^2 Q_i + DM,
	\end{align}
	where $DS$ and $DM$ are the Hessian metrics, given by
	\bea
	DS\!&=&\! \frac{\partial^2 S}{\partial Q_i \partial Q_j} \delta Q_i \delta Q_j +  \frac{2\, \partial^2 S}{\partial Q_i \partial M} \delta Q_i \delta M + \frac{\partial^2 S}{\partial M^2} \delta M^2,\nn\\
	DM\! &=&\! \frac{\partial^2 M}{\partial Q_i \partial Q_j} \delta Q_i \delta Q_j +  \frac{2\, \partial^2 M}{\partial Q_i \partial S} \delta Q_i \delta S + \frac{\partial^2 M}{\partial S^2} \delta S^2.\label{w7}
	\eea
It is straightforward to establish that $DM=\delta T \delta S + \delta \phi_i \delta Q_i=- T DS$ \cite{Liu:2010sz}. Note that in inequality (2), the quantity $\delta^2 A^{KN}$, as given in Eq.~(113) in \cite{Wald1}, is actually the Hessian metric of the area, {\it i.e.}~it is $4DS$.  We now see that the inequality \eqref{w1} can be derived simply by imposing the conditions \cite{Ningbo}
		\begin{equation}
			\delta S = 0\,, \qquad \delta^2 S \geq 0\,, \label{cond}
		\end{equation}
on our black hole thermodynamic sysem. It then follows from eq.~(\ref{w6}), and using $S = A/4$ and $T = \kappa/(2\pi)$, we have the inequality (2).		

The conditions (\ref{cond}) can be derived from requiring $0 \leq \Delta S \equiv \lambda \delta S + \frac{1}{2}\lambda^2 \delta^2 S + \cdots$, which ensures the total change in entropy is nonnegative. When the leading term $\delta S > 0$, it dominates, and we can conclude there is no violation of WCCC at this order. (Alternatively, if $\delta S > 0$, we can redefine $\delta S$ to absorb the higher-order corrections.) To extend our analysis to the next order, it is natural to impose condition \eqref{cond} to maintain the energy condition. This approach has the advantage of avoiding $\delta T$, which appears as $\delta T\delta S$ in the second variation and cannot be determined if the final state after perturbation is not a black hole. Recall that at the first-order, imposing $\delta S > 0$ on the linear variation $M[\lambda]$ (\ref{w5}) leads to inequality \eqref{NEC}, which can also be derived from the NEC \cite{Poisson,Ht1}. For the second-order, applying conditions (\ref{cond}) to the second-order variation of $M[\lambda]$ (\ref{w6}) yields the inequality \eqref{w1}, which can also be obtained when the non-electromagnetic stress-energy tensor satisfies the NEC \cite{Wald2}.

From \eqref{cond}, the second equation of \eqref{w6} becomes an inequality:
		\begin{equation}\label{inequal1}
			\delta^2 M \geq \phi_i \delta^2 Q_i + \delta \phi_i \delta Q_i\,.
		\end{equation}
The extremal limit always saturates this inequality, owing to the fact that $T\delta^2 S$ in \eqref{w6} vanishes. We now examine how this inequality influences the sign of $X_\epsilon$. Since we shall work on the near extremal case, it is convenient to adopt the canonical ensemble with $(T_\epsilon, Q_i)$ as variables.
	The quantity \(X_{\epsilon}\) in \eqref{M1} involves three terms. The first bracket term yields \eqref{e1}. The last term gives
\bea
\Delta M_{\text{ext}}(Q_i) = \lambda \phi_{\text{ext},i} \delta Q_i + \fft{1}{2} \lambda^2 \left( \phi_{\text{ext},i} \delta^2 Q_i +\frac{\partial^2 M_{\text{ext}}}{\partial Q_i \partial Q_j} \delta Q_i \delta Q_j \right).\label{e2}
\eea
	Together with \(\Delta M = \lambda \delta M + \frac{1}{2} \lambda^2 \delta^2 M\) and $\delta M= \phi_i \delta Q_i$ (since $\delta S=0$,)  Eq.~(\ref{M1}) becomes
	\begin{eqnarray}
		X_{\epsilon} &=& \ft12 W\, T_{\epsilon}^2 + \lambda (\phi_i - \phi_{\text{ext},i}) \delta Q_i + \frac{1}{2} \lambda^2 \left( \delta^2 M - \phi_{\text{ext},i} \delta^2 Q_i - \frac{\partial^2 M_{\text{ext}}}{\partial Q_i \partial Q_j} \delta Q_i \delta Q_j \right)\nn\\
		&\ge &\ft12 W\, T_{\epsilon}^2 - \lambda T_\epsilon\,\delta S_{\rm ext} +\ft12 \lambda^2 (\delta \phi_i -\delta \phi_{{\rm ext}, i}) \delta Q_i\,,\label{e3}
	\end{eqnarray}
	where $W$ was defined in \eqref{wdef}. In the last term, we made use of the inequality \eqref{inequal1}. To evaluate this last term further, we notice that $\phi_i=\phi_i(S, Q_j)$ and $\phi_{{\rm ext}, i}= \phi_{{\rm ext}, i} (S_{\rm ext}(Q_j), Q_j)$. Together with requiring $\delta S=0$, we find, up to the zeroth order of $\epsilon$, that
	\begin{align}
	(\delta \phi_i-\delta \phi_{\text{ext,i}})\delta Q_i = - \delta S_{\rm ext}\, \Big(\fft{\partial \phi_i}{\partial S}\Big)_{Q_i;T=0}\, \delta Q_i =-\delta S_{\rm ext}\, \Big(\fft{\partial T}{\partial Q_i}\Big)_{S;T=0} \delta Q_i =
	W^{-1} (\delta S_{\rm ext})^2\,.\label{lambdasqterm}
\end{align}
	In the second equality above, we used the integrability condition of the first law; in the last equality, we used the cyclic differential identity \eqref{cyclicid}. We therefore have an inequality of the SW type, namely
	\be
	X_{\epsilon} \ge \fft12 W \Big( T_\epsilon - \lambda \fft{\delta S_{\rm ext}}{W}\Big)^2.\label{waldnext}
	\ee
	Compared to \eqref{hmain}, we see that Hubeny's type lacks the {\it positive} term \eqref{lambdasqterm} of the $\lambda^2$ order.  In Appendix B we showed the result in Eq. (\ref{waldnext}) is exactly equivalent to (\ref{wald1}) for the KN black hole. Note that in the extreme limit where \(\epsilon=0\) or equivalently \(T_\epsilon = 0\), we have the equal sign, namely
	\be
	X_{\rm ext} = \fft12 \lambda^2 W^{-1} (\delta S_{\rm ext})^2\,,\label{waldnext1}
	\ee
	owing to the fact that the inequality $T \delta^2 S \ge 0$ that was applied to \eqref{w6} is saturated at $T=0$.
	
	\section{On the possibility of WCCC violation}
	
	Using the gedanken experiment and expanding the horizon condition and energy conditions, given by \(\delta  M\) and \(\delta^2 M\) in Eqs.~(\ref{NEC}) and (\ref{w1}), we calculated \(X_{\epsilon}\) and $X_{\rm ext}$ in Eq.~(\ref{waldnext}) for the general near-extreme and extreme black holes, up to the appropriate orders. We see that the sign of $W$ plays a crucial rule in WCCC. It is important to clarify that the sign is an intrinsic property of an extremal black hole, whereas in the testing of the WCCC \eqref{waldnext}, we made no assumption of the final outcome in Eq.~\eqref{M1} after perturbing the black hole, except for applying \eqref{cond}.
	
	In Appendix C, we provide an independent proof of the positivity of \(W\) for spherically-symmetric and static black holes. In Appendix D, we considered two interesting examples: charged Born-Infeld (BI) \cite{BI1,BI2} and Myers-Perry (MP) \cite{MP} black holes in more detail. While the MP black hole is not spherically-symmetric, it respects the WCCC. For general black holes that are specified by conserved mass $M$ and charges $Q_i$, (which include, {\it e.g.~}electromagnetic charges and/or angular momenta,) and have extremal limit, we can deduce from \eqref{e1} that \(W\) is always positive, provided that the extremal mass of the black hole is a local minimum for fixed the charges \(Q_i\) {\it i.e.}
	\be
	M(T_\epsilon,Q_i)-M_{\text{ext}}(Q_i) > 0\,,\quad \leftrightarrow \quad W > 0\,.\label{W>0}
	\ee
Therefore, $W>0$ is closely related to the black hole no-hair theorem \cite{hair1,hair2}, which states that a black hole of given mass, charge and angular momentum is uniquely defined. We must have $M>M_{\rm ext}$; otherwise, two black holes with the same mass and charges could have different temperatures. This is because, for fixed charges, the mass is a function of temperature with $M(T=0)=M_{\rm ext}$. If $M(T)<M_{\rm ext}$ as $T\rightarrow 0$, there should be a minimum mass $M_{\rm min}<M_{\rm ext}$ at certain non-zero temperature, since it is reasonable to assume that black hole mass is unbound above. In this scenario, there would then exist at least two black holes of the same mass slightly above the minimum mass $M_{\rm min}$, but with different temperatures, violating the uniqueness property of black holes. It is also worth pointing out that $W = \lim_{T\rightarrow 0} T^{-1} C_{Q_i}$. Thus the sign of $W$, of an extremal black hole, is the same as the sign of the specific heat capacity $C_{Q_i}$ of the corresponding near-extremal black hole.
	
	Having established that $W>0$, we thus conclude from \eqref{waldnext} and \eqref{waldnext1} that WCCC cannot be violated by the gedanken experiment at the second order. On the other hand, the no-hair theorem is relatively weak, and we might entertain the possibility of $W<0$, even though no such an exotic example is known to exist. In this case, $M_{\rm ext}$ becomes a local maximum and it follows that $X_{\epsilon}$ or $X_{\rm ext}$ should be {\it negative} in order to protect WCCC. Thus WCCC is still protected for extremal black hole, while it could be violated for the near-extremal cases, since \eqref{waldnext} is now insufficient to ensure that $X_{\epsilon}$ is negative. The fact that WCCC is always protected in the extremal case, regardless the sign of $W$, is intriguing.
	
	Our discussion has been focused on the asymptotically-flat spacetimes, since
	by geodesic calculation, one can verify that NEC for an infalling test particle to reach the horizon is equivalent to imposing the second law of thermodynamics \(\delta S>0\) \cite{Zhang,Mt4}. (See Appendix A for the KN black hole example). However, such a connection is lacking when the black holes are asymptotic to (A)dS spacetimes, where the energy and angular momenta of a geodesic particle are less well-defined. Nevertheless, we may still discuss the implication of $\delta S>0$, or more precisely the second-order condition \eqref{cond}, on WCCC for asymptotic (A)dS black holes.
	
	Since our test on WCCC \eqref{waldnext} depends on the sign of $W$, which is a local quantity on the horizon, our conclusion remains the same for asymptotic AdS black holes. However, subtleties arise for asymptotically-dS black holes, since there exists a cosmic horizon $r_c$. In general such a charged and/or rotating black hole has inner, outer and cosmic horizons $(r_-, r_+, r_c)$. We shall not be concerned with the situation of having a double root $r_+\sim r_c$, even though it has $W<0$, since it indicates the collapse of the static regions of universe; it is a case beyond the consideration of WCCC. For extremal or near extremal black holes with $r_+-r_-\sim 0 \ll r_c-r_+$, the conclusion is the same regarding the WCCC. However, it was suggested in \cite{Ningbo} that when a cosmic horizon is involved, one should consider ``total'' entropy $\widetilde S = S_+ + S_c$. As we can see in the Appendix E, the corresponding temperature $\widetilde T$ and electric potential $\widetilde \Phi$ must be
	\be
	\widetilde T=\fft{T_+ T_c}{T_+ + T_c}\,,\qquad \widetilde \Phi_i= \fft{T_c}{T_+ + T_c}\Phi_{+,i} +
	\fft{T_+}{T_+ + T_c}\Phi_{c,i}\,.
	\ee
	Consequently, the first law $dM = \widetilde T d\widetilde S + \widetilde \Phi_i dQ_i$ is satisfied with the same mass and charges. In this case, the condition \eqref{cond} should be replaced with $\delta \widetilde S=0$ and $\delta^2 \widetilde S>0$. Then the conclusion \eqref{waldnext} takes the same form, but with $S$ and $T$ replaced by the corresponding tilded values. In particular, we expect that there is no WCCC violation owing to the fact that $\widetilde W$ is positive.  This ensures that our main conclusion in this paper does not change, whichever entropy definition one takes. It is worth pointing out that in the extremal limit, both definitions of the temperature vanish, namely $T_+=0=\widetilde T$. In Appendix E.2, we illustrate with the KN-dS black hole that
	\be\label{pr}
	\Big(\fft{\partial S_+}{\partial T_+}\Big)_{Q;T_+=0} = \Big(\fft{\partial \widetilde S}{\partial \widetilde T}\Big)_{Q;\widetilde T=0}\,.
	\ee
	However, whether this is generally true or not does not affect our conclusion on the dS black holes.
	
	\section{Conclusion and discussion}
	
	In this paper, we have proposed a novel approach to examine the WCCC in the framework of gedanken experiments, using Eqs.~(\ref{M1}) and (\ref{M2}) for near-extremal and extremal black holes, respectively. This approach allows us to systematically examine the WCCC by providing general formulae for different types discussed in recent decades and enables us to efficiently and clearly solve the contradictions presented in many studies, such as \cite{Wald,Wald1,Wald2, Sanjar1,Sanjar2, Hubney, Ht1,Ht2,Ht3,Yang1,Yang2, Zhang,Mt1,Mt2, Mt3,Mt4,Mt5,Sanjar0,Ningbo}. We categorized well-known types of WCCC violations as Hubeny's type \cite{Hubney}, mixed type \cite{Zhang}, and SW  type \cite{Wald1}, and derived general formulae in Eqs.~(\ref{hmain}), (\ref{main0}), and (\ref{waldnext}), respectively for each case, emphasizing the importance of appropriate order expansions and avoiding mixing orders in the near-extremal and extremal limits.
	
	We demonstrated that our general formula for each case provided a concise summary of the results obtained in various papers that claimed to have counterexamples. Focusing on the SW type, we first verified the Wald inequality (\ref{w1}) by adopting a new approach of applying the condition \eqref{cond} \cite{Ningbo}. We found that the general testing of WCCC depends on the sign of $W$ \eqref{wdef} for both near-extremal and extremal black holes, where if $W > 0$, WCCC is protected, while for $W < 0$, it indicates potential violation near the extreme case. For the extreme case, WCCC is protected regardless the sign of $W$.
	
	The condition $W>0$ is in fact an intrinsic property of an extremal black hole, whereas it also appears in the testing of WCCC (\ref{waldnext}) after perturbing the black hole. We illustrated that the $W>0$ condition is equivalent to $M(T_\epsilon,Q_i)>M_{\text{ext}}(Q_i)$. This condition is also closely related to the black hole no-hair theorem. We also independently proved $W>0$ for the general spherically-symmetric and static extremal black holes.
	
	For asymptotically flat spacetimes, the NEC for test particles reaching the event horizon is linked to $\delta S>0$ or can be obtained by geodesic calculations, while this connection is absent for asymptotically (A)dS spacetimes,
where the energy and angular momenta of a geodesic test particle is not well defined. If we assume that the condition \eqref{cond} continues to hold when cosmological constant is involved, we showed, for the same reason of $W>0$, there is no violation of WCCC for (A)dS black holes. However, in the case of dS spacetime, one might consider new thermodynamic system involving both outer and cosmic horizons, with the entropy given by $\widetilde S=S_+ + S_c$. We showed that there could be no violation of WCCC again, for $\widetilde W>0$.
	
	Our results provide a clear and general framework for analyzing the WCCC in various black hole solutions, reducing the testing of WCCC to only examining the sign of $W$. This approach helps to avoid misunderstandings of the conceptual and technical issues, ensuring the validity of the calculations. Future work could extend these general formulas to more complex spacetimes and higher-order gravities to examine the possibility of WCCC violation in asymptotically flat or AdS/dS spacetimes. Furthermore, quantum and back-reaction effects could be included in the calculation process. Additionally, exploring the role of different energy conditions and their implications on WCCC in various black hole scenarios or theories could provide deeper insights into understanding WCCC in gedanken experiments.
	
	\section{Acknowledgement}
	
	This work was supported in part by NSFC (National Natural Science Foundation of China) Grants No. 11935009 and No. 12375052.
	

\clearpage

\appendix

\section{Hubeny and mixed types}

According to \cite{Wald,Zhang, Mt4}, for the KN BH, the outer horizon is given by the greater root of \(\Delta = r^2 + a^2 + Q^2 - 2Mr = 0\), which is \(r_+ = M + \sqrt{M^2 - Q^2 - a^2}\). Solving both \(\Delta = 0\) and \(\partial_r \Delta = 0\) simultaneously leads to
\begin{equation}\label{extrM}
	M_{\text{ext}} = r_{\text{ext}} = \frac{\sqrt{Q^2 + \sqrt{4 J^2 + Q^4}}}{\sqrt{2}}.
\end{equation}
The thermodynamic properties of the KN BH are as
\[
\phi_a = \frac{a}{r_+^2 + a^2}, \qquad T = \frac{r_+ - M}{2\pi (r_+^2 + a^2)}, \qquad \phi_Q = \frac{r_+ Q}{r_+^2 + a^2}, \qquad S = \pi (r_+^2 + a^2).
\]
The first law reads $
\delta M = T \delta S + \phi_a \delta J + \phi_Q \delta Q$,
where \(\delta\) denotes variations in the KN BH's parameters. For an infalling particle with energy \(E\), angular momentum \(L\), and charge \(q\), the conserved energy and angular momentum are
\[
E = -(mu_\mu + q A_\mu) \left(\frac{\partial}{\partial t}\right)^\mu, \quad L = (mu_\mu + q A_\mu) \left(\frac{\partial}{\partial \phi}\right)^\mu,
\]
where \(u^\mu\) is its four-velocity. Given \(E\) and \(L\), we may eliminate \(\dot{\phi}\) in the above relations and obtain \(\dot{t}\) as
\[
\dot{t} = \frac{g_{\phi\phi}(E + q A_t) + g_{t\phi}(L - q A_{\phi})}{m(g_{t\phi}^2 - g_{\phi\phi} g_{tt})}.
\]
To ensure a future-pointed velocity, \(\dot{t} > 0\) is necessary, which implies that the numerator of the above relation must be positive. This provides a lower bound for \(E\), which can be evaluated at the outer horizon of the BH, yielding
\begin{equation}
	E \geq \frac{aL + qQr_+}{a^2 + r_+^2}.
	\label{1^}
\end{equation}
If the particle crosses through the outer horizon, then the mass, charge, and angular momentum of the BH change. Meanwhile, some energy of the particle is lost by radiation. Thus, according to the conservation laws,
\begin{align}
	\delta M < E, \quad \delta J = L, \quad \delta Q = q.
	\label{x1}
\end{align}
Here we assume that the initial spin of the particle aligns with the axis of symmetry of the BH. By symmetry, it remains parallel to the BH's axis after absorption, ensuring that the BH remains axisymmetric and that any radiation from the particle does not carry away any angular momentum. Considering \(\delta S > 0\) and \(\delta M < E\), this equation is equivalent to the inequality obtained from the first law as
\begin{equation}
	E > \delta M \geq \phi_i \delta Q_i = \frac{a \delta J + Qr_+ \delta Q}{a^2 + r_+^2}.
	\label{1*}
\end{equation}
Hence, if the black hole is slightly perturbed by a probe, the change in the BH's parameters should satisfy (\ref{1*}), or equivalently (\ref{1^}).
Since we are interested in destroying the horizon (\(M^2 < Q^2 + a^2\)) by throwing a particle into it, we must have
\begin{equation}
	(\delta M + M)^2 - (Q + q)^2 - \left(\frac{aM + L}{M + \delta M}\right)^2 < 0\,,
	\label{3*}
\end{equation}
or equivalently
\begin{equation}\label{do}
	(E + M)^2 - (Q + q)^2 - \left(\frac{aM + L}{M + E}\right)^2 < 0\,.
\end{equation}

\subsection{Near Extreme KN BH: Hubeny's Type}

According to \cite{Ht3}, following the Hubeny method for the near-extreme KN BH, it is suggested that WCCC protection requires \(\text{eq} > 0\), where we have
\begin{equation}\label{do0}
	\text{eq} \equiv E - \frac{M_{\text{ext}} Q q + a L}{M_{\text{ext}}^2 + a^2} + \frac{1}{2} \frac{M_{\text{ext}}^3}{M_{\text{ext}}^2 + a^2} \left(\frac{\delta}{M_{\text{ext}}}\right)^2 ,
\end{equation}
where we have \(\delta^2 = M^2 - a^2 - Q^2\). To calculate \(E\), we have \(E > \delta M \geq \phi_i \delta Q_i\). Then expanding \(\phi_i\) near the extreme case, where \(r = M_{\text{ext}} (1 + \epsilon)\), leads to
\begin{equation}\label{do2}
	E > \phi_{\epsilon,i} \delta Q_i = \frac{M_{\text{ext}} Q q + a L}{M_{\text{ext}}^2 + a^2} - \frac{M_{\text{ext}} (2 M_{\text{ext}} a L + q Q^3)}{(a^2 + M_{\text{ext}}^2)^2} \epsilon + \ldots \, .
\end{equation}
Substituting (\ref{do2}) into (\ref{do0}) gives
\begin{equation}\label{do1}
	eq > \frac{1}{2} \frac{M_{\text{ext}}^3}{M_{\text{ext}}^2 + a^2} \epsilon^2 -  \frac{M_{\text{ext}} (2 M_{\text{ext}} a L + q Q^3)}{(a^2 + M_{\text{ext}}^2)^2} \epsilon ,
\end{equation}
in which \(\epsilon = \delta / M_{\text{ext}}\), \(a = J / M_{\text{ext}}\) are used.

To see the possibility of WCCC violation, one needs to check whether the inequality (\ref{do1}) can be satisfied or not. One can also obtain the same result as presented in Eq.~(\ref{do1}) by using the formula we provided in Eq.~(\ref{hmain}). To do so for Hubeny's type, we have
\begin{equation}\label{do3}
	W\,  T_{\epsilon}^2 = \frac{M_{\text{ext}}^3}{M_{\text{ext}}^2 + a^2} \epsilon^2, \quad \quad T_{\epsilon} \delta S_{\text{ext}} = \frac{M_{\text{ext}} (2 M_{\text{ext}} a L + q Q^3)}{(a^2 + M_{\text{ext}}^2)^2} \epsilon.
\end{equation}
Substituting the above equations into Eq.~(\ref{hmain}) yields the same result as presented in Eq.~(\ref{do1}), where \(\delta Q = q\) and \(\delta J = L\) are used.

\subsection{Extreme KN BH: mixed type}

Now, expanding relation (\ref{do}) up to the first order in \(E/M\), \(\delta Q/Q\), and \(\delta J/(aM)\) at the extremal limit \(M_{\text{ext}}^2 - a^2 - Q^2 = 0\) gives
\begin{equation}
	E < \frac{a \delta J - QM_{\text{ext}} \delta Q}{M_{\text{ext}}^2 + a^2}.
\end{equation}
Thus, by expanding the destroying horizon condition up to the first order and comparing it with Eq.~(\ref{1*}), we find no violation of the WCCC (this was first mentioned by Wald \cite{Wald}). In \cite{Zhang}, Eq.~(\ref{do}) is expanded up to the second order, and using Eq.~(\ref{1*}) at the extreme limit, the following relation is obtained for WCCC violation (see Eq.~(41) in \cite{Zhang})
\begin{equation}\label{xp}
	P= \frac{2 a^2 M_{\text{ext}}^2 (3 M_{\text{ext}}^2 - a^2)}{(a^2 + M_{\text{ext}}^2)^3} q^2 + \frac{M^2 (-3 M_{\text{ext}}^2 + a^2)}{(a^2 + M_{\text{ext}}^2)^3} L^2 - \frac{2 a M_{\text{ext}} Q (3 M_{\text{ext}}^2 - a^2)}{(a^2 + M_{\text{ext}}^2)^3} qL > 0,
\end{equation}
where \(L = \delta J\) and \(q = \delta Q\). It is shown that small values of \(L\) always lead to a violation of WCCC. For more examples, please see the papers cited \cite{Mt1, Mt2, Mt3, Mt4, Mt5}. In \cite{Mt4}, avoiding any expansion, it is shown that violation of CCC is still possible with fine-tuning of the energy, charge, and angular momentum of a test particle or scalar field for both extreme and near-extreme KN BHs.

Now we want to use the provided formula in Eq. (\ref{main0}) for the KN BH, to show it leads to the same results in (\ref{xp}), where \( \delta Q_i = q_i \). Substituting \( M_{\text{ext}} \) from Eq. (\ref{extrM}) into (\ref{main0}) leads to
\begin{align}\label{X}
	X_{\text{ext,mix}} = & \frac{\sqrt{Q^2 + \sqrt{4 J^2 + Q^4}} \left(4 J^2 - Q^4 + Q^2 \sqrt{4 J^2 + Q^4}\right)}{2 \sqrt{2} (4 J^2 + Q^4)^{3/2}} q^2 \nonumber \\
	& + \frac{(2 Q^2 - \sqrt{4 J^2 + Q^4}) \sqrt{Q^2 + \sqrt{4 J^2 + Q^4}}}{2 \sqrt{2} (4 J^2 + Q^4)^{3/2}} L^2 \nonumber \\
	& - \frac{\sqrt{2} J \left(4 J^2 Q + 3 Q^3 \left(Q^2 + \sqrt{4 J^2 + Q^4}\right)\right)}{(4 J^2 + Q^4)^{3/2} (Q^2 + \sqrt{4 J^2 + Q^4})^{3/2}} qL.
\end{align}
We can write the above equation in terms of \( M_{\text{ext}} \) and \( a \), in the same form as Eq. (\ref{xp}). Substituting
\begin{equation}
	Q = \sqrt{M_{\text{ext}}^2 - \frac{J^2}{M_{\text{ext}}^2}}, \quad J = a M_{\text{ext}}
\end{equation}
into (\ref{X}), we have
\begin{equation}\label{xpn}
	X_{\text{ext,mix}} = \frac{a^2 M_{\text{ext}} (3 M_{\text{ext}}^2 - a^2)}{(a^2 + M_{\text{ext}}^2)^3} q^2 + \frac{M_{\text{ext}} (-3 M_{\text{ext}}^2 + a^2)}{2 (a^2 + M_{\text{ext}}^2)^3} L^2 - \frac{a Q (3 M_{\text{ext}}^2 - a^2)}{(a^2 + M_{\text{ext}}^2)^3} q L
\end{equation}
where \(X_{\text{ext,mix}} = P / (2M_{\text{ext}}\)). It is expected that the final result will differ from that proposed in \cite{Zhang} with a coefficient \(2 M_{\text{ext}} \), as this paper employs Eq. (\ref{3*}) for the destroying horizon condition. Therefore, in the mixed type of WCCC one can use the general formula in Eq. (\ref{main0}) instead of going through complicated processes in different examples. Many examples have been considered using the mixed type framework, such as \cite{Mt1, Mt2, Mt3, Mt4, Mt5}, where their results can be concluded by a straightforward calculation as presented in (\ref{main0}).

\section{Verifying Eq.~(\ref{waldnext}) for KN BH}

To verify the result obtained in Eq. (\ref{waldnext}), considering the near-extremal KN black hole we have
\begin{align}
	&\epsilon = r_+/M - 1 = \frac{\sqrt{M^2-Q^2-J^2/M^2}}{M}\,,\nn\\
	&T_{\epsilon} = \frac{\epsilon M^3}{2 \pi (J^2 + (1+\epsilon)^2 M^4)} = \frac{\epsilon M^3}{2 \pi (J^2 + M^4)} + O(\epsilon)^2\,,\nn\\
	&\delta S_{\rm ext}=\frac{2 \pi \left(2 \, J \delta J M^2 + Q \delta Q (M^4-J^2) \right)}{J^2 + M^4}\,,\qquad W = \frac{4 \pi^2 \left(J^2 + M^4\right)}{M}\,.
\end{align}
Then one can verify Wald's result as
\begin{equation}\label{wald0}
	X_{\epsilon} \ge \frac{1}{2} W \left( T_\epsilon - \lambda \frac{\delta S_{\rm ext}}{W}\right)^2 = \frac{1}{2} \frac{M^3}{J^2 + M^4} \left(M \epsilon - \lambda \frac{2 J \delta J M^2 + Q \delta Q (M^4-J^2)}{M (J^2 + M^4)}  \right)^2,
\end{equation}
Note that at the above order, the mass \(M\) should be taken to be the extremal value \(M_{\rm ext}\). In Eq. (\ref{wald0}), the coefficient is different from Wald's calculation in Eq. (\ref{wald1}). Regarding this point, one can expand \(M(\lambda) = M(T(\lambda), Q_i(\lambda))\) near \(M_{\text{ext}}(\lambda) = M_{\text{ext}}(Q_i(\lambda))\) in Eq. (\ref{w8}) as
\begin{equation}\label{f1}
	f(\lambda) = f(M(\lambda), M_{\text{ext}}(\lambda)) = f(M_{\text{ext}}(\lambda), M_{\text{ext}}(\lambda)) + \frac{\partial f}{\partial M} \bigg|_{M=M_{\text{ext}}} (M(\lambda) - M_{\text{ext}}(\lambda)) + \dots,
\end{equation}	
where \(f(M_{\text{ext}}(\lambda), M_{\text{ext}}(\lambda)) = 0\). On the other hand, one can write
\begin{align}
	M(\lambda) - M_{\text{ext}}(\lambda) &= M(0) + \lambda M'(0) + \frac{1}{2} M''(0) - M_{\text{ext}}(0) + \lambda M_{\text{ext}}'(0) + \frac{1}{2} M_{\text{ext}}''(0) \notag\\
	&= M(T_\epsilon, Q^i) - M_{\text{ext}}(Q^i) + \Delta M - \Delta M_{\text{ext}} = X_{\epsilon},
\end{align}
so that we have
\begin{equation}\label{f2}
	f(\lambda) = \frac{\partial f}{\partial M} \bigg|_{M=M_{\text{ext}}} X_{\epsilon} + \cdots\,,\qquad
	\frac{\partial f}{\partial M} \bigg|_{M=M_{\text{ext}}} = \frac{2 \left(J^2 + M_{\text{ext}}^4\right)}{M_{\text{ext}}^3}\,,
\end{equation}
where $M_{\rm ext}$ was given in \eqref{wald1}. Therefore, the result in Eq.~(\ref{wald0}), obtained from \(X_{\epsilon}\), can be written exactly the same as Wald's result in Eq.~(\ref{wald1}).

\section{$W>0$ for spherically-symmetric and static black holes}

Here, we provide an independent proof of the positivity of \(W\), defined as \eqref{wdef}, for spherically-symmetric and static black holes
\be
ds^2 = - e^{2\chi} f dt^2 + \fft{dr^2}{f} + r^2 d\Omega_2^2\,,
\ee
The \(f\) and \(\chi\) are functions of
\be
f=f(r,M, Q_i)\,, \qquad \chi=\chi(r,M, Q_i)\,.
\ee
The horizon radius $r_+$ satisfies $f(r_+, M, Q_i)=0$, which implies that $M=M(r_+, Q_i)$. The temperature is then given by
\be
T= \fft{e^{\chi} f'(r,M(r_+),Q_i)}{4\pi}\Big|_{r=r_+}\,.
\ee
A prime denotes a derivative with respect to \(r\). Thus we have
\be
\Big(\fft{\partial S}{\partial T}\Big)_{Q_i} = \frac{8 \pi^2 r_+\, e^{-\chi}}{f'' + 2 \pi T (2e^{-\chi} \chi' + r_+ \partial_M f') - 8\pi^2 r_+ T^2 \partial_M e^{-\chi}}\bigg|_{r=r_+,\,M=M(r_+)}\,.
\ee
Here we used the thermodynamic identity that $\partial_{r_+} M= 2\pi r_+ T$.
The above is valid for the the general non-extremal black hole. In the extremal limit, we have
\be
T \sim f'(r_+) = 0\,, \qquad f''(r_+) > 0\,.
\ee
Thus we have
\be
W=\Big(\fft{\partial S}{\partial T}\Big)_{Q_i;T=0} = \frac{8 \pi^2 r_+ e^{-\chi(r_+)}}{f''(r_+)} > 0\,.
\ee

\section{Explicit examples}

In this section, we are going to calculate \(W\), defined by \eqref{wdef}, for MP and Charged BI BHs to show there is no possibility of WCCC violation for these BHs.

\subsection{Charged BI BH}

The spherically-symmetric and asymptotically-flat solution in Einstein-Born-Infeld (EBI) theory was given as \cite{BI1,BI2}
\bea
f &=& 1 - \fft{2M}{r} + \fft{r^2 - \sqrt{r^4 + 2\beta Q^2}}{3\beta}
+ \fft{4Q^2}{3r^2} {}_2F_1[\ft14, \ft12; \ft54, -2\beta Q^2/r^4]\,, \nn\\
a &=& -\frac{Q}{r} \, _2F_1[\ft{1}{4}, \ft{1}{2}; \ft{5}{4}, -2\beta Q^2/r^4]\,A = a(r) dt\,.
\eea
The solution reduces to the standard RN solution in the \(\beta \rightarrow 0\) limit. The matter energy-momentum tensor \(T^{\mu}_\nu = \text{diag}\{-\rho, p_r, p_T\}\) of the BI field is given by
\be
\rho = -p_r =  \fft{\sqrt{2 \beta Q^{2} + r^{4}} - r^{2}}{\beta r^{2}}\,, \qquad
\rho + p_T =  \fft{2 Q^{2}}{r^{2} \sqrt{2 \beta Q^{2} + r^{4}}}\,.
\ee
Therefore, we have also
\be
\rho - P_{T} = \fft{(\sqrt{2 \beta Q^{2} + r^{4}} - r^{2})^{2}}{\beta r^{2} \sqrt{\beta Q^{2} + r^{4}}}\,, \qquad
\rho + p_{r} + 2p_{T} =  \fft{2(\sqrt{2 \beta Q^{2} + r^{4}} - r^{2})}{\beta \sqrt{2 \beta Q^{2} + r^{4}}}\,.
\ee
This implies that the NEC is satisfied and for \(\beta \ge 0\), both strong and dominant energy conditions are satisfied. For \(\beta < 0\), the dominant energy condition is violated, but the strong and weak energy conditions remain satisfied.

All the thermodynamic quantities as functions of \((Q, r_+)\):
\bea
M &=& \frac{1}{6 \beta r_+} \Big(4 \beta Q^2 \, _2F_1\left(\frac{1}{4}, \frac{1}{2}; \frac{5}{4}, -\frac{2 Q^2 \beta}{r_+^4}\right) - r_+^2 \left(\sqrt{2 \beta Q^2 + r_+^4} - 3 \beta \right) + r_+^4 \Big)\,, \nn\\
T &=& \frac{\beta - \sqrt{2 \beta Q^2 + r_+^4} + r_+^2}{4 \pi \beta r_+}\,, \qquad
S = \pi r_+^2\,, \qquad \Phi = \fft{Q}{r_+} \, _2F_1\left(\frac{1}{4}, \frac{1}{2}; \frac{5}{4}, -\frac{2 \beta Q^2}{r_+^4}\right).
\eea
It is easy to verify that \(dM = TdS + \Phi dQ\).

In the extremal limit, we denote the horizon as \(r_0\) and we have
\be
r_0 = \sqrt{Q_{\rm ext}^2 - \ft12 \beta}\,, \qquad \rightarrow \qquad \beta \le 2 Q^2\,.
\ee
This condition is easier to be satisfied by a negative \(\beta\). The existence of zero temperature, however, requires that \(\beta + 2Q_{\rm ext}^2 \ge 0\). Thus we must have for extremal black holes
\be\label{beta1}
-2Q_{\rm ext}^2 \le \beta \le 2Q_{\rm ext}^2\,, \qquad \hbox{or} \qquad  |\beta| \le 2 Q_{\rm ext}^2\,.
\ee

We find
\be
W= \Big(\fft{\partial S}{\partial T}\Big)_{Q_i; T=0} = \frac{\pi^2}{2} (4 Q^2 + 2 \beta) \sqrt{4 Q^2 - 2 \beta}.
\ee
Thus it appears to have no violation in the valid range of \(\beta\) in Eq. (\ref{beta1}).

\subsection{Myers-Perry BH}

The spherically-symmetric and asymptotically-flat solution of the MP BH in 5 dimensions is given by \cite{MP}
\begin{equation}\label{MP}
	f(r) = \frac{(r^2 + a^2)(r^2 + b^2)}{r^2} - \mu\,, \qquad \mu = \frac{8 M}{3 \pi}\,, \qquad a = \frac{4 J_a}{\pi \mu}\,, \qquad
	b = \frac{4 J_b}{\pi \mu}\,.
\end{equation}
The MP BH satisfies all the standard energy conditions in its exterior (vacuum) region, similar to other vacuum solutions like the Schwarzschild or Kerr BH. The thermodynamic quantities as a function of \((a, b, r_+)\) can be written as
\begin{equation}\label{MP1}
	T = \frac{(r_+^2 f'(r_+))}{4 \pi (r_+^2 + a^2) (r_+^2 + b^2)}\,, \quad
	S = \frac{\pi^2 (r_+^2 + a^2) (r_+^2 + b^2)}{2 r_+}\,, \quad
	\phi_a = \frac{a}{r_+^2 + a^2}\,, \quad
	\phi_b = \frac{b}{r_+^2 + b^2}\,,
\end{equation}
where the first law can be verified as
$
\delta M = T \delta S + \phi_a \delta J_a + \phi_b \delta J_b.
$
For the extreme \(\mu\) and \(r_+\) we also have
\begin{equation}\label{MP2}
	\mu_{\text{ext}} = (a + b)^2; \quad \quad r_{\text{ext}} = \sqrt{ab}.
\end{equation}
Finally, one can find
\bea
W = \Big(\fft{\partial S}{\partial T}\Big)_{Q_i; T=0} = \frac{\pi^3}{4} (a + b)^4,
\eea
which obviously shows that there is no violation of WCCC.

\section{Asymptotically-dS black holes}

In the following, we study the thermodynamic properties of asymptotically-dS black holes. In particular, we consider the thermodynamics with the entropy as the total sum of those associated the outer and cosmic horizons $r_+<r_c$, each of which satisfies the respective first law:
\be
dM = T_+ dS_+ + \Phi_{+,i} dQ_i\,,\qquad  dM = T_c dS_c + \Phi_{c,i} dQ_i\,.
\ee
It is clear that \(T_c<0\) and \(T_+>0\), with both \(S_+\) and \(S_c\) being positive. Now one can write the two differential equations as
\be
\fft{1}{T_+} dM =  dS_+ + \fft{\Phi_{+,i}}{T_+} dQ_i\,,\qquad \fft{1}{T_c} dM =  dS_c + \fft{\Phi_{c,i}}{T_c} dQ_i\,,
\ee
which gives
\be
\Big(\fft{1}{T_+} + \fft{1}{T_c}\Big) dM =  d(S_+ +S_-) + \Big(\fft{\Phi_+}{T_+} + \fft{\Phi_c}{T_{c,i}}\Big) dQ_i\,.
\ee
This leads to
\be
\fft{1}{\widetilde T}=\fft{1}{T_+} + \fft{1}{T_c}\,,\qquad \widetilde S=S_+ + S_c\,,\qquad \widetilde \Phi_i = \Big(\fft{\Phi_{+,i}}{T_+} + \fft{\Phi_{c,i}}{T_c}\Big) \widetilde T\,,
\ee
which satisfy the first law of the same mass and charge, namely
\be
dM = \widetilde T d\widetilde S +\widetilde \Phi_i dQ_i\,.
\ee
It is important to note that in the extremal $T_+=0$ limit, we have $\widetilde T=0$. We can thus re-derive the inequality \eqref{waldnext} based on the above first law and the inequality condition \eqref{cond}, all with $(T, S, \Phi)$ replaced by the corresponding tilded values. The outcome then depends on the sign of $\widetilde W$
\be
\widetilde W = \Big(\fft{\partial\widetilde S}{\partial \widetilde T}\Big)_{Q_i; \widetilde T=0}\,.
\ee
It is clear that analogous argument leads to $\widetilde W>0$.  In what follows, we consider RN-dS and KN-dS black holes where $\widetilde W = W$.

\subsection{RN-dS BH}

Here we present the RN-dS black hole as a concrete example. The metric profile function $f$
\be
f = -\ft13 \Lambda r^2 + 1 - \fft{2M}{r^2} + \fft{Q^2}{r^2}\,.
\ee
For appropriate mass $M$ and charge $Q$, there are three positive roots, and we label them as \(r_-\le r_+ < r_c\). Then the fourth root must be negative, namely \(-(r_- + r_+ + r_c)\), and one can write
\be
f = -\frac{\left(r - r_-\right) \left(r - r_+\right) \left(r - r_c\right) \left(r + r_c + r_- + r_+\right)}{r^2 \left(r_- r_c + r_c^2 + r_+ r_c + r_-^2 + r_+ r_- + r_+^2\right)}\,.
\ee
The mass, charge, and the cosmological constant can be expressed in terms of the three horizon radii as
\bea
M &=& \frac{\left(r_- + r_+\right) \left(r_c + r_-\right) \left(r_c + r_+\right)}{2 \left(r_c^2 + \left(r_- + r_+\right) r_c + r_-^2 + r_+ r_- + r_+^2\right)}\,,\nn\\
Q^2 &=& \frac{r_- r_+ r_c \left(r_c + r_- + r_+\right)}{r_- \left(r_c + r_+\right) + r_c^2 + r_+ r_c + r_-^2 + r_+^2}\,,\nn\\
\Lambda &=& \frac{3}{r_c^2 + \left(r_- + r_+\right) r_c + r_-^2 + r_+ r_- + r_+^2}\,.
\eea
It is encouraging to see that all the above three quantities are positive.  We further have
\bea
&&T_- = \frac{\left(r_- - r_+\right) \left(r_c - r_-\right) \left(r_c + 2 r_- + r_+\right)}{4 \pi r_-^2 \left(r_c^2 + \left(r_- + r_+\right) r_c + r_-^2 + r_+ r_- + r_+^2\right)} < 0\,,\nn\\
&&T_+ = \frac{\left(r_+ - r_-\right) \left(r_c - r_+\right) \left(r_c + r_- + 2 r_+\right)}{4 \pi r_+^2 \left(r_c^2 + \left(r_- + r_+\right) r_c + r_-^2 + r_+ r_- + r_+^2\right)} > 0\,,\nn\\
&&T_c = -\frac{\left(r_c - r_-\right) \left(r_c - r_+\right) \left(2 r_c + r_- + r_+\right)}{4 \pi r_c^2 \left(r_c^2 + \left(r_- + r_+\right) r_c + r_-^2 + r_+ r_- + r_+^2\right)} < 0\,,\\
&&\widetilde{T} = \frac{\left(r_+ - r_-\right) \left(r_c - r_-\right) \left(r_c + r_- + 2 r_+\right) \left(2 r_c + r_- + r_+\right)}{4 \pi \left(r_c+r_+\right) \left(\left(r_--r_+\right) r_c+r_- \left(r_-+r_+\right)\right) \left(r_c^2+\left(r_-+r_+\right) r_c+r_-^2+r_+ r_-+r_+^2\right)}>0\,.\nn
\eea
The requirement that \(\Lambda\) and \(Q\) be constants implies
\bea
dr_- &=& -\frac{r_- \left(r_+ - r_c\right) \left(r_c + r_- + 2 r_+\right)}{r_+ \left(r_- - r_c\right) \left(r_c + 2 r_- + r_+\right)}\,dr_+\,,\nn\\
dr_c &=& \frac{\left(r_- - r_+\right) r_c \left(r_c + r_- + 2 r_+\right)}{r_+ \left(r_c - r_-\right) \left(2 r_c + r_- + r_+\right)}\, dr_+\,.
\eea
We can now calculate
\be
\Big(\fft{\partial S_+}{\partial T_+}\Big)_{Q;T_+ =0} = \frac{4 \pi^2 r_+^3 \left(2 r_+ r_c + r_c^2 + 3 r_+^2\right)}{\left(r_c - r_+\right) \left(r_c + 3 r_+\right)}
= \Big(\fft{\partial \widetilde{S}}{\partial \widetilde{T}}\Big)_{Q;\widetilde{T} = 0}\,.
\ee
Note that \(T_+ = 0 \leftrightarrow \widetilde{T} = 0\), achieved by setting \(r_- = r_+\).
For \(r_c > r_+\), the above two (same) quantities are both positive, indicating that the RN-dS black holes cannot violate WCCC.

\subsection{KN-dS BH}

We now consider the more general KN-dS black hole, where the horizon is determined by \cite{KNads}
\be
\Delta = (r^2 + a^2) \left(1 - \ft13 \Lambda r^2\right) - 2m r + q^2\,.
\ee
The root of \(\Delta\) is denoted as \(r_i\), with \(i = -, +, c\), associated with inner, outer, and cosmic horizons. The temperature and entropies are then given by
\be
T_i = \frac{r_i^2 - a^2 - q^2}{4 \pi r_i \left(a^2 + r_i^2\right)} - \frac{\Lambda r_i \left(a^2 + 3 r_i^2\right)}{12 \pi \left(a^2 + r_i^2\right)}\,,
\qquad S_i = \fft{\pi (r_i^2 + a^2)}{\Xi}\,,
\ee
where
\be
\Xi = 1 + \ft13 \Lambda a^2\,.
\ee
We can express the three parameters \((m, q, \Lambda)\) in terms of \((r_-, r_+, r_c)\), given by
\bea
m &=& \frac{\left(r_- + r_+\right) \left(r_c + r_-\right) \left(r_c + r_+\right)}{2 \left(a^2 + r_c^2 + \left(r_- + r_+\right) r_c + r_-^2 + r_+^2 + r_- r_+\right)}\,,\nn\\
q^2 &=& \frac{r_- r_+ r_c \left(r_c + r_- + r_+\right)}{a^2 + r_c^2 + \left(r_- + r_+\right) r_c + r_-^2 + r_+^2 + r_- r_+} - a^2\,,\nn\\
\Lambda &=& \frac{3}{a^2 + r_c^2 + \left(r_- + r_+\right) r_c + r_-^2 + r_+^2 + r_- r_+}\,.
\eea
Now \(\Delta\) is simply given by
\be
\Delta = -\frac{\left(r - r_-\right) \left(r - r_+\right) \left(r - r_c\right) \left(r_c + r + r_- + r_+\right)}{a^2 + r_c^2 + \left(r_- + r_+\right) r_c + r_-^2 + r_+^2 + r_- r_+}\,.
\ee
The mass, charge, and angular momentum are
\be
M = \fft{m}{\Xi^2}\,,\qquad J = \fft{m a}{\Xi^2}\,,\qquad Q = \fft{q}{\Xi}\,.
\ee
Requiring \(\delta J = \delta Q = \delta \Lambda = 0\) allows us to solve for \(\delta r_-, \delta r_c\) and \(\delta a\) in terms of \(\delta r_+\). In particular, we find that
\be
\fft{\partial r_-}{\partial r_+}\Big|_{Q, J; T=0} = -1\,, \qquad
\fft{\partial r_c}{\partial r_+}\Big|_{Q, J; T=0} = 0\,, \qquad
\fft{\partial a}{\partial r_+}\Big|_{Q, J; T=0} = 0\,.
\ee
We therefore again have
\be
\Big(\fft{\partial S_+}{\partial T_+}\Big)_{Q, J; T_+=0} = \frac{4 \pi^2 r_+ \left(a^2 + r_+^2\right) \left(a^2 + r_c^2 + 2 r_+ r_c + 3 r_+^2\right)^2}{\left(r_c - r_+\right) \left(r_c + 3 r_+\right) \left(2 a^2 + r_c^2 + 2 r_+ r_c + 3 r_+^2\right)}
= \Big(\fft{\partial \widetilde{S}}{\partial \widetilde{T}}\Big)_{Q,J;\widetilde{T} = 0}\,.
\ee	
			
\end{document}